\documentclass[aps,twocolumn,superscriptaddress]{revtex4}

\usepackage{graphicx}

\def\ket#1{\left| #1\right>}

\begin{document}


\title{Decoherence dynamics of a qubit coupled to a quantum two-level system}

\author{S. Ashhab}
\affiliation{Frontier Research System, The Institute of Physical
and Chemical Research (RIKEN), Wako-shi, Saitama, Japan}
\author{J. R. Johansson}
\affiliation{Frontier Research System, The Institute of Physical
and Chemical Research (RIKEN), Wako-shi, Saitama, Japan}
\author{Franco Nori}
\affiliation{Frontier Research System, The Institute of Physical
and Chemical Research (RIKEN), Wako-shi, Saitama, Japan}
\affiliation{Center for Theoretical Physics, CSCS, Department of
Physics, University of Michigan, Ann Arbor, Michigan, USA}


%
%
%


\date{\today}

\begin{abstract}
We study the decoherence dynamics of a qubit coupled to a quantum
two-level system (TLS) in addition to its weak coupling to a
background environment. We analyze the different regimes of
behaviour that arise as the values of the different parameters are
varied. We classify those regimes as two weak-coupling regimes,
which differ by the relation between the qubit and TLS decoherence
times, and a strong-coupling one. We also find analytic
expressions describing the decoherence rates in the weak-coupling
regimes, and we verify numerically that those expressions have a
rather wide range of validity. Along with obtaining the
above-mentioned results, we address the questions of qubit-TLS
entanglement and the additivity of multiple TLS contributions. We
also discuss the transition from weak to strong coupling as the
parameters are varied, and we numerically determine the location
of the boundary between the two regimes.
\end{abstract}


\maketitle


\section{Introduction}

There have been remarkable advances in the quest to build a
superconductor-based quantum information processor in recent years
\cite{You,Nakamura1,Nakamura2,Vion,Yu,Martinis1,Ciorescu,Pashkin,Yamamoto,
Simmonds,Cooper,Astafiev,Ithier,Bertet}. Coherent oscillations
have been observed in systems of single qubits and two interacting
qubits \cite{Nakamura1,Vion,Yu,Pashkin}. In order to achieve the
desirable power of a functioning quantum computer, one would need
to perform a large number of quantum gate operations of at least
hundreds of qubits. One of the main obstacles to achieving that
goal, however, are the relatively short decoherence times in these
macroscopic systems (note that even a single superconductor-based
qubit can be considered a macroscopic system). Therefore, there
has been increasing experimental and theoretical activity aimed at
understanding the sources and mechanisms of decoherence of such
systems in recent years
\cite{Simmonds,Cooper,Astafiev,Ithier,Bertet,Paladino,Galperin2,Ku,Shnirman,Faoro,Galperin,Saikin}.

The environment causing decoherence of the qubit is composed of a
large number of microscopic elements. There is a large wealth of
theoretical work on the so-called spin-boson model \cite{Leggett},
which models the environment as a large set of harmonic
oscillators, to describe the environment of a solid-state system.
However, recent experimental results suggest the existence of
quantum two-level systems (TLSs) that are strongly coupled to the
qubit \cite{Simmonds,Cooper,Martinis2}. Furthermore, it is well
known that the qubit decoherence dynamics can depend on the exact
nature of the noise causing that decoherence. For example, an
environment composed of a large number of TLSs that are all weakly
coupled to the qubit will generally cause non-Markovian
decoherence dynamics in the qubit (see, e.g., \cite{Paladino}).
Note that the mechanism of qubit-TLS coupling depends on the
physical nature of the qubit and TLS. The exact mechanisms are
presently unknown.

The effect on the qubit of an environment consisting of weakly
coupled TLSs with short decoherence times is rather well
understood. As was presented in Ref. \cite{Shnirman}, one takes
the correlation functions of the TLS dynamics in the frequency
domain, multiplies each one with a factor describing the qubit-TLS
coupling strength, and adds up the contributions of all the TLSs
to obtain the effective noise that is felt by the qubit. We shall
refer to that approach as the traditional weak-coupling
approximation. In this paper we shall study the more general case
where no {\it a priori} assumptions are made about the TLS
parameters. We shall identify the criteria under which the
traditional weak-coupling approximation is valid. We shall also
derive more general expressions that have a wider range of
validity, as will be discussed below. Furthermore, we study the
criteria under which our weak-coupling results break down, and the
TLS cannot be easily factored out of the problem. It is worth
noting here that we shall not attempt to theoretically reproduce
the results of a given experiment. Although we find potentially
measurable deviations from the predictions of previous work, we
are mainly interested in answering some questions related to the
currently incomplete understanding of the effects of a TLS, or
environment of TLSs, on the qubit decoherence dynamics.

Since we shall consider in some detail the case of a weakly
coupled TLS, and we shall use numerical calculations as part of
our analysis, one may ask why we do not simulate the decoherence
dynamics of a qubit coupled to a large number of such TLSs.
Alternatively, one may ask why we separate one particular TLS from
the rest of the environment. Focussing on one TLS has the
advantage that we can obtain analytic results describing the
contribution of that TLS to the qubit decoherence. That analysis
can be more helpful in building an intuitive understanding of the
effects of an environment composed of a large number of TLSs than
a more sophisticated simulation of an environment composed of,
say, twenty TLSs. The main purpose of using the numerical
simulations in this work is to study the conditions of validity of
our analytically obtained results.

The present paper is organized as follows: in Sec. II we introduce
the model system and the Hamiltonian that describes it. In Sec.
III we describe the theoretical approach that we shall use in our
analysis. In Sec. IV we use a perturbative calculation to derive
analytic expressions for the relaxation and dephasing rates of the
qubit in the weak-coupling regime and compare them with those of
the traditional weak-coupling approximation. In Sec. V we
numerically analyze the qubit decoherence dynamics in the
different possible regimes. We also address a number of questions
related to the intuitive understanding of the problem, including
those of qubit-TLS entanglement and the case of two TLSs. In Sec.
VI we discuss the question of the boundary between the weak and
strong coupling regimes, and we perform numerical calculations to
determine the location of that boundary. We finally conclude our
discussion in Sec. VII.

\section{Model system}

We consider a qubit that is coupled to a quantum TLS. We take the
qubit and the TLS to be coupled to their own (uncorrelated)
environments that would cause decoherence even if the qubit and
TLS are not coupled to each other. We shall be interested in the
corrections to the qubit decoherence dynamics induced by the TLS.
The Hamiltonian of the system is given by:

\noindent
\begin{equation}
\hat{H} = \hat{H}_{\rm q} + \hat{H}_{\rm TLS} + \hat{H}_{\rm I} +
\hat{H}_{\rm Env},
\end{equation}

\noindent where $\hat{H}_{\rm q}$ and $\hat{H}_{\rm TLS}$ are the
qubit and TLS Hamiltonians, respectively, $\hat{H}_I$ describes
the coupling between the qubit and the TLS, and $\hat{H}_{\rm
Env}$ describes all the degrees of freedom in the environment and
their coupling to the qubit and the TLS. The qubit Hamiltonian is
given by:

\noindent
\begin{equation}
\hat{H}_{\rm q} = -\, \frac{\Delta_{\rm q}}{2} \hat{\sigma}_x^{\rm
(q)} - \frac{\epsilon_{\rm q}}{2} \hat{\sigma}_z^{\rm (q)},
\end{equation}

\noindent where $\Delta_{\rm q}$ and $\epsilon_{\rm q}$ are the
adjustable control parameters of the qubit, and
$\hat{\sigma}_{\alpha}^{\rm (q)}$ are the Pauli spin matrices of
the qubit. For example, for the charge qubit in Ref.
\cite{Nakamura1}, $\Delta_{\rm q}$ and $\epsilon_{\rm q}$ are the
energy scales associated with tunnelling and charging,
respectively. Similarly, the TLS Hamiltonian is given by:

\noindent
\begin{equation}
\hat{H}_{\rm TLS} = -\, \frac{\Delta_{\rm TLS}}{2}
\hat{\sigma}_x^{\rm (TLS)} - \frac{\epsilon_{\rm TLS}}{2}
\hat{\sigma}_z^{\rm (TLS)}.
\end{equation}

\noindent The energy splitting between the two quantum states of
each system, in the absence of coupling between them, is then
given by:

\noindent
\begin{equation}
E_{\alpha} = \sqrt{{\Delta_{\alpha}}^2+{\epsilon_{\alpha}}^2},
\end{equation}

\noindent where the index $\alpha$ refers to either qubit or TLS.
For future purposes, let us also define the angles
$\theta_{\alpha}$ by the criterion
\begin{equation}
\tan\theta_\alpha \equiv \frac{\Delta_\alpha}{\epsilon_\alpha}.
\end{equation}
We take the interaction Hamiltonian between the qubit and the TLS
to be of the form:

\noindent
\begin{equation}
\hat{H}_I = - \frac{\lambda}{2} \hat{\sigma}_z^{\rm (q)} \otimes
\hat{\sigma}_z^{\rm (TLS)},
\end{equation}

\noindent where $\lambda$ is the coupling strength between the
qubit and the TLS. Note that the minus sign in $\hat{H}_I$ is
simply a matter of convention, since $\lambda$ can be either
positive or negative. It is worth mentioning here that the
applicability of this form of interaction is not as limited as it
might appear at first sight. Any interaction Hamiltonian that is a
product of a qubit observable (i.e. any Hermitian $2 \times 2$
matrix) and a TLS observable can be recast in the above form with
a simple basis transformation.

We assume that all the terms in $\hat{H}_{\rm Env}$ are small
enough that its effect on the dynamics of the qubit+TLS system can
be treated within the framework of the Markovian Bloch-Redfield
master equation approach. It is well known that the effect of
certain types of environments cannot be described using that
approach, e.g. those containing $1/f$ low-frequency noise.
However, there remain a number of unanswered questions about the
problem of a qubit experiencing $1/f$ noise. For example, it was
shown in Ref. \cite{Galperin2} that the decoherence dynamics can
depend on the specific physical model used to describe the
environment. Therefore, treating that case would make it more
difficult to extract results that are directly related to the
phenomenon we are studying, namely the effect of a single quantum
TLS on the qubit decoherence dynamics. We therefore do not
consider that case.

Depending on the physical nature of the system, the coupling of
the qubit and the TLS to their environments is described by
specific qubit and TLS operators. In principle one must use those
particular operators in analyzing the problem at hand. However,
since we shall present our results in terms of the background
decoherence rates, which are defined as the relaxation and
dephasing rates in the absence of qubit-TLS coupling, the choice
of system-environment interaction operators should not affect any
of our results. In fact, in our numerical calculations below we
have used a number of different possibilities and verified that
the results remain unchanged, provided that the background
decoherence rates are kept constant. Furthermore, the
background-noise power spectrum affects the results only through
the background decoherence rates. Note that we shall not discuss
explicitly the temperature dependence of the background
decoherence rates. It should be kept in mind, however, that the
background dephasing rates generally have a strong temperature
dependence in current experiments on superconducting qubits.

It is worth noting that, since we are considering a
quantum-mechanical TLS, the model and the intermediate algebra
that we use are essentially identical to those used in some
previous work studying two coupled qubits
\cite{Thorwart,Storcz,You2}. However, as opposed to being a second
qubit, a TLS is an uncontrollable and inaccessible part of the
system. Therefore, in interpreting the results, we only consider
quantities related to the qubit dynamics.

\section{Theoretical analysis: Master equation}

As mentioned above, we take one particular element of the
environment, namely the TLS, and do not make any {\it a priori}
assumptions about its decoherence times or the strength of its
coupling to the qubit. We assume that the coupling of the qubit to
its own environment and that of the TLS to its own environment are
weak enough that a Markovian master equation approach provides a
good description of the dynamics. The combined qubit+TLS system
has four quantum states. The quantity that we consider is
therefore the $4 \times 4$ density matrix describing that combined
system. We follow the standard procedure to write the
Bloch-Redfield master equation as (see e.g. Ref.
\cite{Cohen_Tannoudji}):

\noindent
\begin{equation}
\dot{\rho}_{ab} = -i \omega_{ab}\; \rho_{ab} + \sum_{cd}
R_{abcd}\; \rho_{cd}, \label{eq:MasterEquation}
\end{equation}

\noindent where the dummy indices $a,b,c$ and $d$ run over the
four quantum states, $\omega_{ab} \equiv (E_a-E_b)/\hbar$, $E_i$
is the energy of the quantum state labelled by $i$, and the
coefficients $R_{abcd}$ are given by:

\noindent
\begin{widetext}
\begin{eqnarray}
R_{abcd} & = & - \int_0^{\infty} dt \sum_{\alpha={\rm q,TLS}}
\Bigg\{ g_{\alpha}(t) \left[ \delta_{bd} \sum_{n} \langle a |
\hat{\sigma}_z^{(\alpha)} | n \rangle \langle n |
\hat{\sigma}_z^{(\alpha)} | c \rangle e^{i\omega_{cn}t} + \langle
a | \hat{\sigma}_z^{(\alpha)} | c \rangle \langle d |
\hat{\sigma}_z^{(\alpha)} |
b \rangle e^{i\omega_{ca}t} \right] \nonumber \\
& & \hspace{1cm} + g_{\alpha}(-t) \left[ \delta_{ac} \sum_{n}
\langle d | \hat{\sigma}_z^{(\alpha)} | n \rangle \langle n |
\hat{\sigma}_z^{(\alpha)} | b \rangle e^{i\omega_{nd}t} + \langle
a | \hat{\sigma}_z^{(\alpha)} | c \rangle \langle d |
\hat{\sigma}_z^{(\alpha)} | b \rangle e^{i\omega_{bd}t} \right]
\Bigg\}
\end{eqnarray}
\end{widetext}
\noindent
\begin{eqnarray}
g_{\alpha}(t) & = & \int_{-\infty}^{\infty} d\omega
S_{\alpha}(\omega) e^{-i\omega t},
\end{eqnarray}

\noindent where $S_{\alpha}(\omega)$ is the background-noise power
spectrum. In calculating $R_{abcd}$ we neglect the imaginary
parts, which renormalize the energy splittings of the qubit and
TLS, and we assume that those corrections are already taken into
account in our initial Hamiltonian. We do not use any secular
approximation to simplify the tensor $R_{abcd}$ any further. One
of the main reasons for avoiding the secular approximation is that
we shall consider cases where the coupling strength between the
qubit and the TLS is very small, which results in almost
degenerate quantum states, a situation that cannot be treated
using, for example, the form of the secular approximation given in
Ref. \cite{Cohen_Tannoudji}.

Once we solve Eq. (\ref{eq:MasterEquation}) and find the dynamics
of the combined system, we can trace out the TLS degree of freedom
to find the dynamics of the reduced $2 \times 2$ density matrix
describing the qubit alone. From that dynamics we can infer the
effect of the TLS on the qubit decoherence and, whenever the decay
can be fit well by exponential functions, extract the qubit
dephasing and relaxation rates.

\section{Analytic results for the weak-coupling limit}

We first consider a case that can be treated analytically, namely
that of a strongly-dissipative weakly-coupled TLS. That is exactly
the case where the traditional weak-coupling approximation is
expected to work. Here we perform a perturbative calculation on
Eq. (\ref{eq:MasterEquation}) where the coupling strength
$\lambda$ is treated as a small parameter in comparison with the
decoherence times in the problem. We shall discuss the differences
between the predictions of the two approaches in this section, and
we shall show in Sec. V that our results have a wider range of
validity than the traditional weak-coupling approximation.

In the first calculation of this section, we consider the
zero-temperature case. If we take the limit $\lambda \rightarrow
0$ and look for exponentially decaying solutions of Eq.
(\ref{eq:MasterEquation}) with rates that approach the unperturbed
relaxation and dephasing rates $\Gamma_1^{\rm (q)}$ and
$\Gamma_2^{\rm (q)}$, we find the following approximate
expressions for the leading-order corrections:

\noindent
\begin{widetext}
\parbox{16cm}{
\begin{eqnarray*}
\delta \Gamma_1^{\rm (q)} & \approx & \frac{1}{2} \lambda^2 \sin^2
\theta_{\rm q} \sin^2 \theta_{\rm TLS} \frac{\Gamma_2^{\rm (TLS)}
+ \Gamma_2^{\rm (q)} - \Gamma_1^{\rm (q)} }{ \left(\Gamma_2^{\rm
(TLS)}+\Gamma_2^{\rm (q)}-\Gamma_1^{\rm (q)}\right)^2 + \left(
E_{q} - E_{\rm TLS} \right)^2}
\\
\delta \Gamma_2^{\rm (q)} & \approx & \frac{1}{4} \lambda^2 \sin^2
\theta_{\rm q} \sin^2 \theta_{\rm TLS} \frac{\Gamma_2^{\rm
(TLS)}-\Gamma_2^{\rm (q)} }{ \left(\Gamma_2^{\rm
(TLS)}-\Gamma_2^{\rm (q)}\right)^2 + \left( E_{q} - E_{\rm TLS}
\right)^2},
\end{eqnarray*}
}
\hfill
\parbox{1cm}{\begin{equation} \label{eq:Pert_theory_rates} \end{equation}}

\noindent which are expected to apply very well when
$\Gamma_2^{\rm (TLS)}+\Gamma_2^{\rm (q)} \gg \Gamma_1^{\rm (q)}$.
The above expressions can be compared with those given in Ref.
\cite{Shnirman}:

\noindent
\parbox{16cm}{
\begin{eqnarray*}
\delta \Gamma_1^{\rm (q)} & \approx & \frac{1}{2} \lambda^2 \sin^2
\theta_{\rm q} \sin^2 \theta_{\rm TLS} \frac{\Gamma_2^{\rm
(TLS)}}{ {\Gamma_2^{\rm (TLS)}}^2 + \left(E_{\rm q} - E_{\rm TLS}
\right)^2}
\\
\delta \Gamma_2^{\rm (q)} & \approx & \frac{1}{2} \delta \Gamma_1
+ \frac{\lambda^2 \cos^2 \theta_{\rm q} \cos^2 \theta_{\rm TLS} }{
\Gamma_1^{\rm (TLS)}} {\rm sech}^2 \left( \frac{E_{\rm
TLS}}{2k_BT} \right).
\end{eqnarray*}
} \hfill
\parbox{1cm}{
\begin{equation}
\label{eq:Semiclassical_rates}
\end{equation}
}
\end{widetext}

\noindent The two approaches agree in the limit where they are
both expected to apply very well, namely when the decoherence
times of the TLS are much shorter than those of the qubit [note
that we are taking $T=0$ in Eq. (\ref{eq:Pert_theory_rates})]. Our
results can therefore be considered a generalization of those of
the traditional weak-coupling approximation. We shall discuss the
range of validity of our results in Sec. V.C.1.

We now turn to the finite temperature case. In addition to
treating $\lambda$ as a small parameter, one can also perform a
perturbative calculation to obtain the temperature dependence of
the decoherence rates in the low temperature limit. In the general
case where the qubit and TLS energy splittings are different and
no assumption is made about the relation between qubit and TLS
decoherence rates, the algebra is rather complicated, and the
resulting expressions contain a large number of terms. Therefore
we only present the results in the case where $E_{\rm q}=E_{\rm
TLS}\equiv E$, which is the case that we shall focus on in Sec. V.
In that case we find the additional corrections to the relaxation
and dephasing rates to be given by:

\noindent
\begin{widetext}
\parbox{16cm}{
\begin{eqnarray*}
\delta \Gamma_{1,T}^{\rm (q)} & = & 0
\\
\delta \Gamma_{2,T}^{\rm (q)} & = & \lambda^2 e^{-E/k_BT} \left(
\frac{ 4 \cos^2 \theta_{\rm q} \cos^2 \theta_{\rm TLS} }{
\Gamma_1^{\rm (TLS)}} - \frac{ \sin^2 \theta_{\rm q} \sin^2
\theta_{\rm TLS} \Gamma_1^{\rm (q)} }{ \left(\Gamma_2^{\rm
(TLS)}-\Gamma_2^{\rm (q)}\right) \left(\Gamma_2^{\rm
(TLS)}-\Gamma_2^{\rm (q)} + \Gamma_1^{\rm (q)} \right)} \right).
\end{eqnarray*}
} \hfill
\parbox{1cm}{
\begin{equation}
\label{eq:emptylbl}
\end{equation}
\vspace{0.5cm} }
\end{widetext}

\noindent Note that the first term inside the parentheses agrees
with the expression given in Eq. (\ref{eq:Semiclassical_rates})
for the TLS contribution to the dephasing rate. If the decoherence
times of the TLS are much shorter than those of the qubit, the
second term is negligible. A similar situation occurs when $E_{\rm
q} \neq E_{\rm TLS}$, i.e., all the terms can be neglected except
for the one given in Eq. (\ref{eq:Semiclassical_rates}).

\section{Numerical solution of the master equation}

We now turn to the task of numerically analyzing the effect of the
TLS on the qubit with various choices of parameters. We analyze
any given case by first solving Eq. (\ref{eq:MasterEquation}) to
find the density matrix of the combined qubit+TLS system as a
function of time. We then trace out the TLS degree of freedom to
find the (time-dependent) density matrix of the qubit alone, which
is perhaps most easily visualized as a curve in the Bloch sphere
\cite{Bloch_sphere}. We then use the Hamiltonian of the qubit
including the mean-field correction contributed by the TLS as a
pole of reference in the Bloch sphere, from which we can extract
the dephasing and relaxation dynamics of the qubit. In other
words, we transform the qubit density matrix into the qubit energy
eigenbasis, such that the diagonal matrix elements describe
relaxation dynamics and the off-diagonal matrix elements describe
dephasing dynamics. The relaxation rate is then defined as the
rate of change of the diagonal matrix elements divided by their
distance from the equilibrium value. The dephasing rate is defined
similarly using the off-diagonal matrix elements
\cite{Fedichkin1}.

Since our main goal is to analyze the different possible types of
behaviour in the qubit dynamics, we have to identify the relevant
parameters that determine the different behaviour regimes. As
discussed in Sec. II, the energy scales in the problem are the
qubit and TLS energy splittings, their background decoherence
rates (which are related to the environment noise power spectrum),
the qubit-TLS coupling strength and temperature. Note that if the
difference between the two energy splittings is substantially
larger than the coupling strength, the effect of the TLS on the
qubit dynamics diminishes rapidly. The above statement is
particularly true regarding the relaxation dynamics. We therefore
consider only the case where the two energy splittings are equal,
i.e. $E_{\rm q}=E_{\rm TLS}$. In other words, we take the TLS to
be on resonance with the qubit. Furthermore, we take the energy
splitting, which is the largest energy scale in the problem, to be
much larger than all other energy scales, such that its exact
value does not affect any of our results. We take the temperature
to be much smaller than $E_{\rm q}$, so that environment-assisted
excitation processes can be neglected. We are therefore left with
the background decoherence rates and the coupling strength as free
parameters that we can vary in order to study the different
possible types of behaviour in the qubit dynamics.

\subsection{Weak-coupling regimes}

Although the discussion of the criterion that distinguishes
between the weak and strong-coupling regimes is deferred to Sec.
VI, we separate the results of this section according to that
criterion. We start with the weak-coupling regimes. In Figs. 1 and
2, we show, respectively, the relaxation and dephasing rates of
the qubit as functions of time for three different sets of
parameters differing by the relation between the qubit and TLS
decoherence rates, maintaining the relation
$\Gamma_2^{(\alpha)}=2\Gamma_1^{(\alpha)}$, where
$\Gamma_1^{(\alpha)}$ and $\Gamma_2^{(\alpha)}$ are the background
relaxation and dephasing rates, i.e. those obtained in the case
$\lambda=0$, and the index $\alpha$ refers to qubit and TLS. The
case $\lambda=0$ is trivial, and we only show it as a point of
reference to demonstrate the changes that occur in the case
$\lambda \neq 0$. All the curves shown in Figs. 1 and 2 agree very
well with the formulae that will be given below.

\begin{figure*}[ht]
\begin{minipage}[b]{5.5cm}
\includegraphics[width=5.3cm]{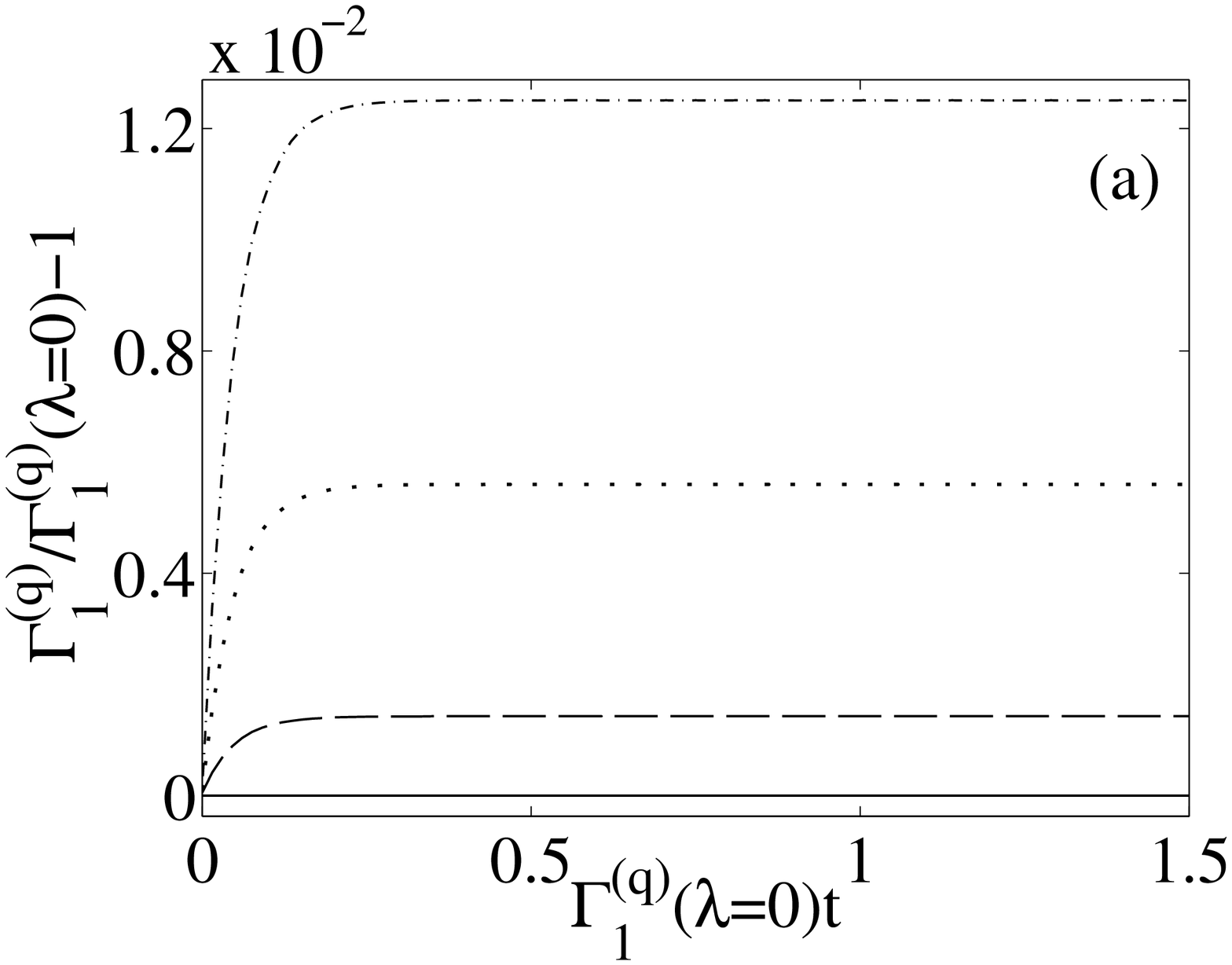}
\end{minipage}
\begin{minipage}[b]{5.5cm}
\includegraphics[width=5.05cm]{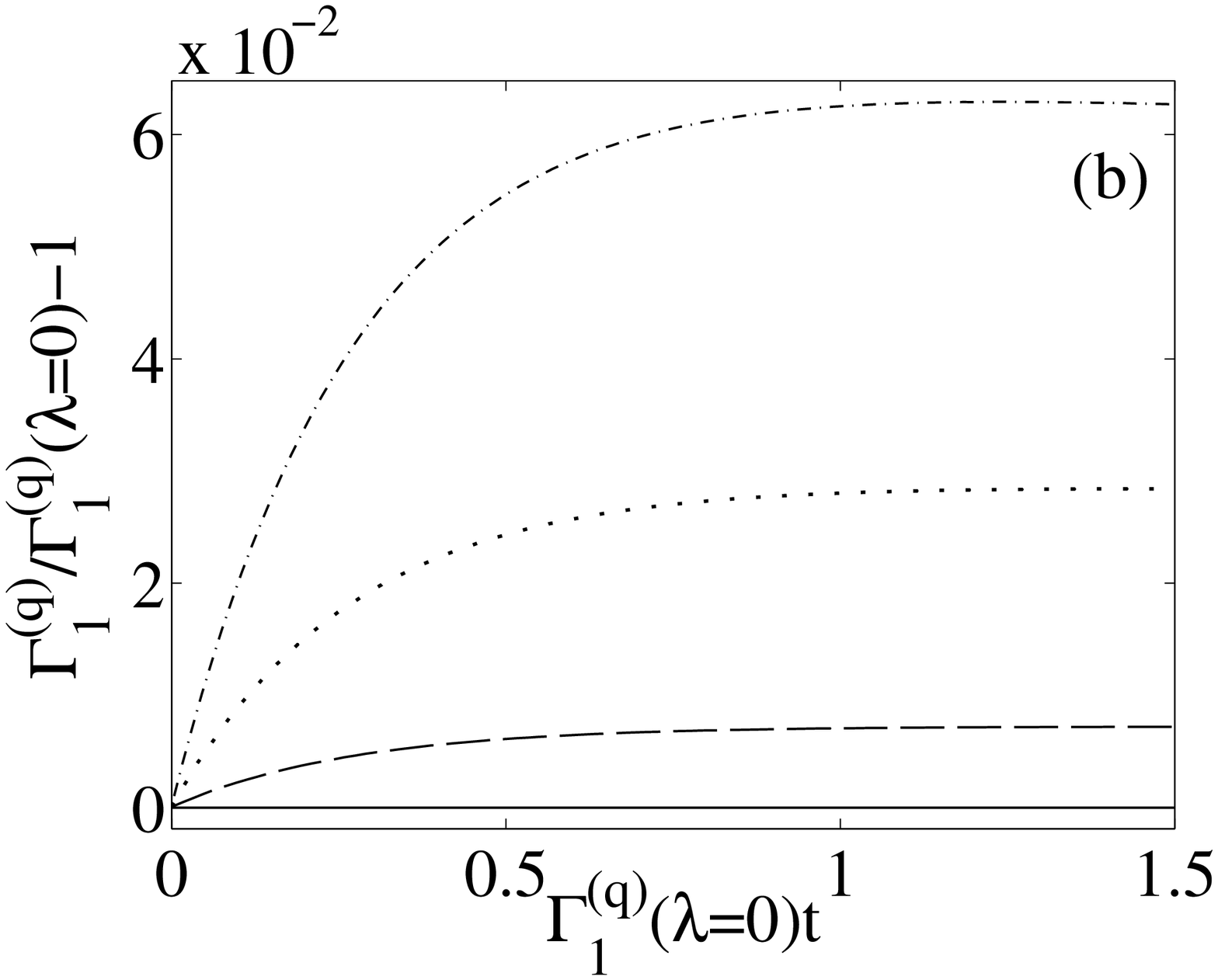}
\end{minipage}
\begin{minipage}[b]{5.6cm}
\includegraphics[width=5.3cm]{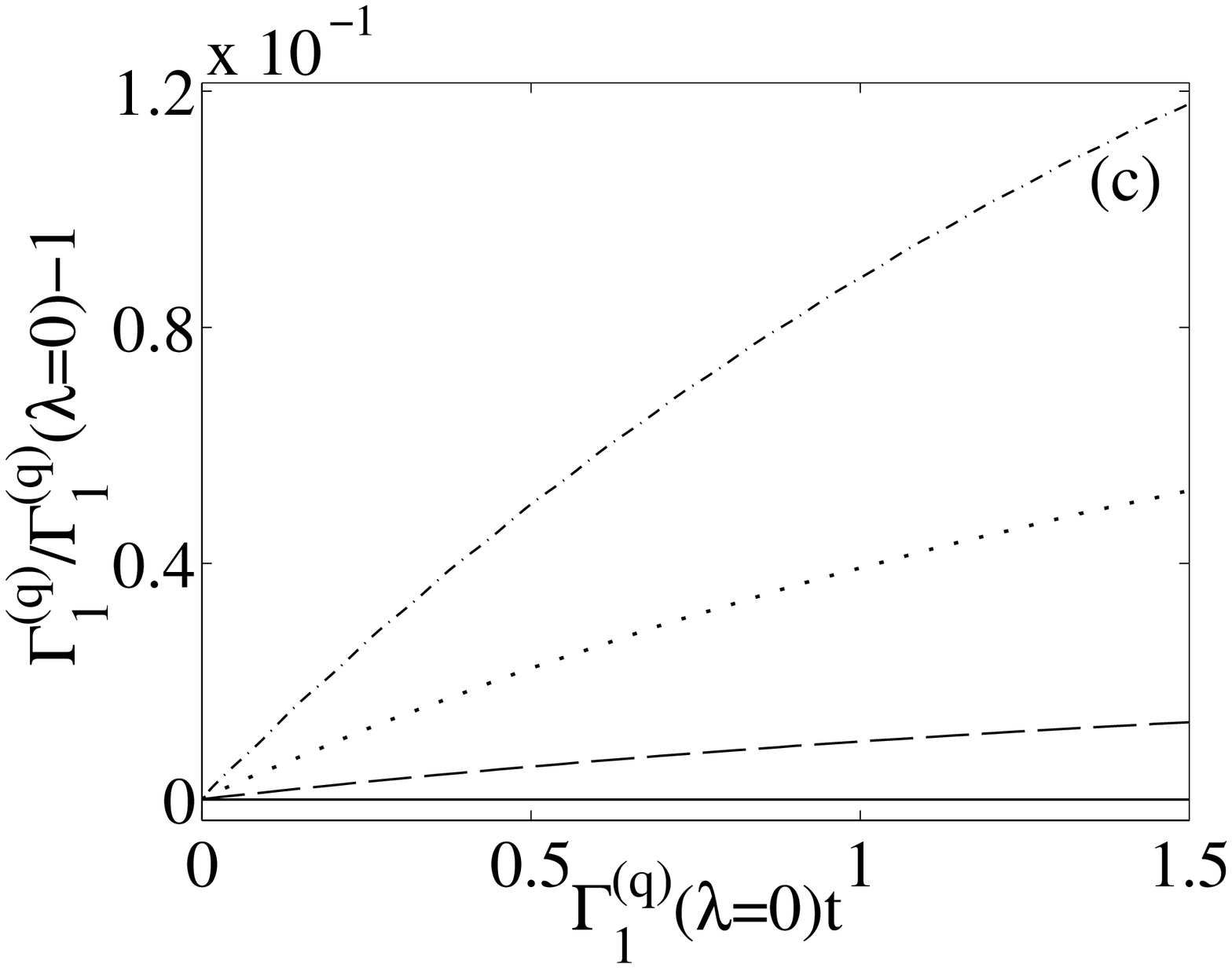}
\end{minipage}
\caption{Relative corrections to qubit relaxation rate as a
function of scaled time in the case of (a) strongly, (b)
moderately and (c) weakly-dissipative TLS. The ratio
$\Gamma_1^{\rm (TLS)}/\Gamma_1^{\rm (q)}$ is 10 in (a), 1.5 in (b)
and 0.1 in (c). The solid, dashed, dotted and dash-dotted lines
correspond to $\lambda/\Gamma_1^{\rm (q)}=0$, 0.3, 0.6 and 0.9,
respectively. $\theta_{\rm q} = \pi/3$ and $\theta_{\rm
TLS}=3\pi/8$.}
\end{figure*}
\begin{figure*}[ht]
\begin{minipage}[b]{5.5cm}
\includegraphics[width=5.1cm]{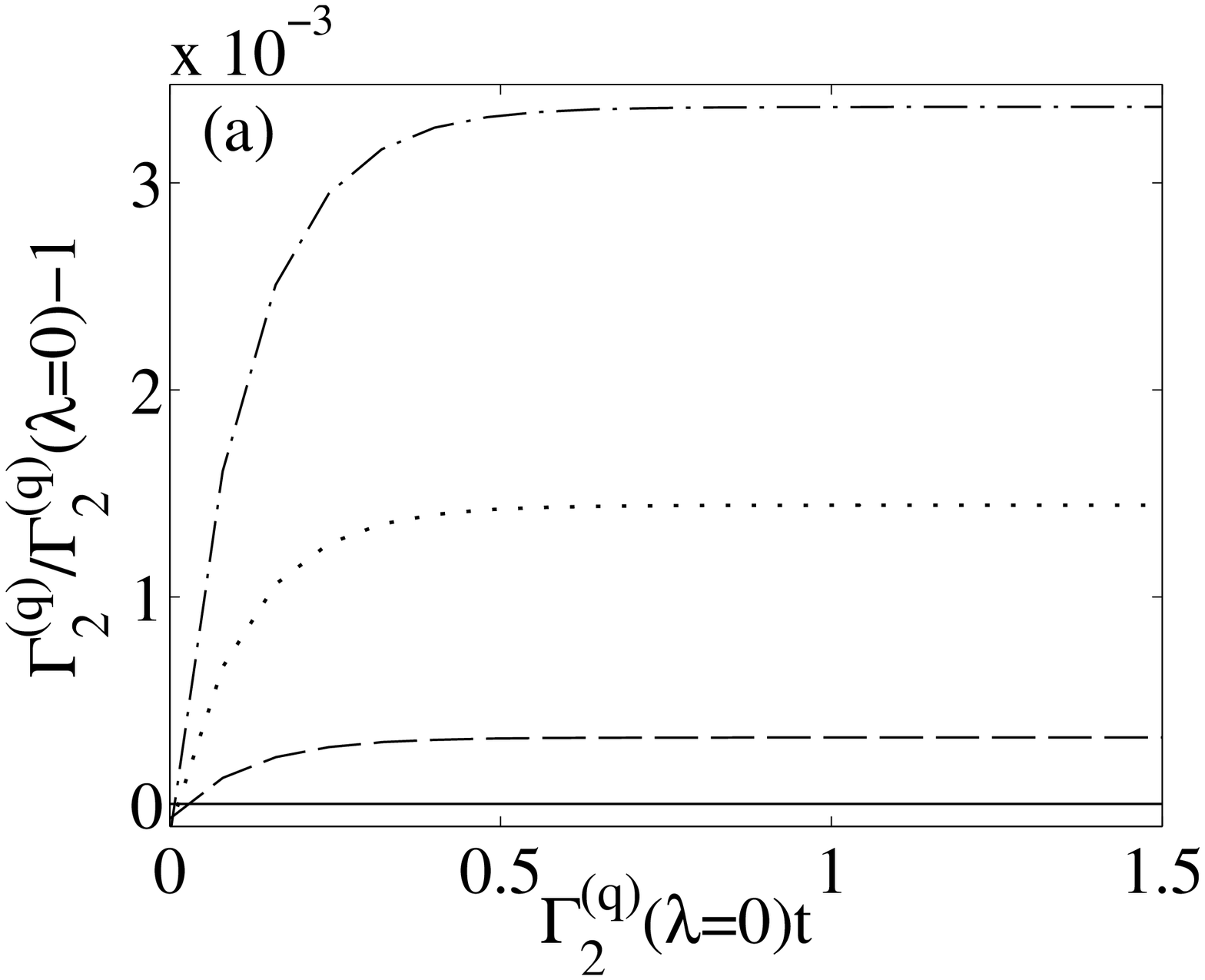}
\end{minipage}
\begin{minipage}[b]{5.6cm}
\includegraphics[width=5.1cm]{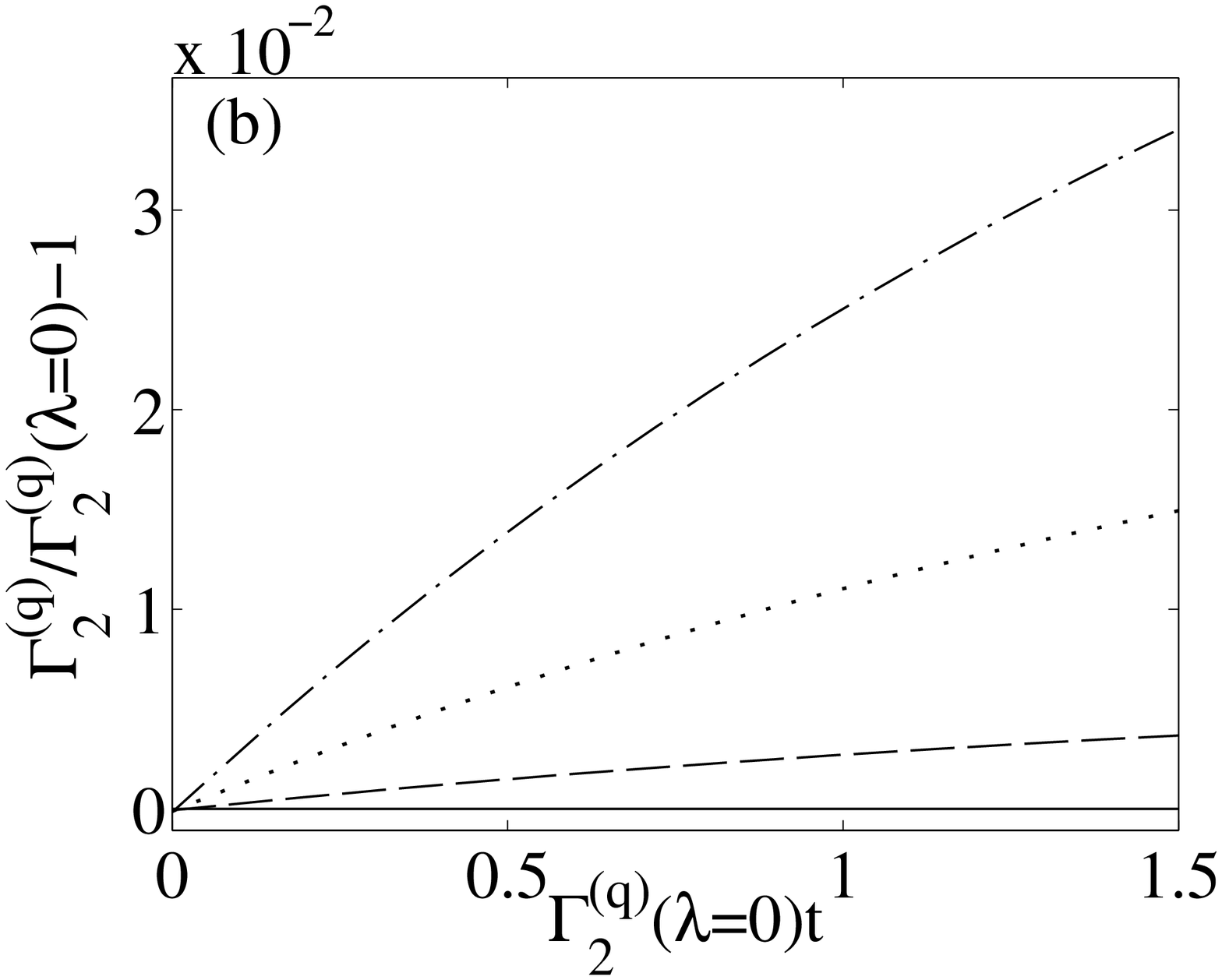}
\end{minipage}
\begin{minipage}[b]{5.5cm}
\includegraphics[width=5.1cm]{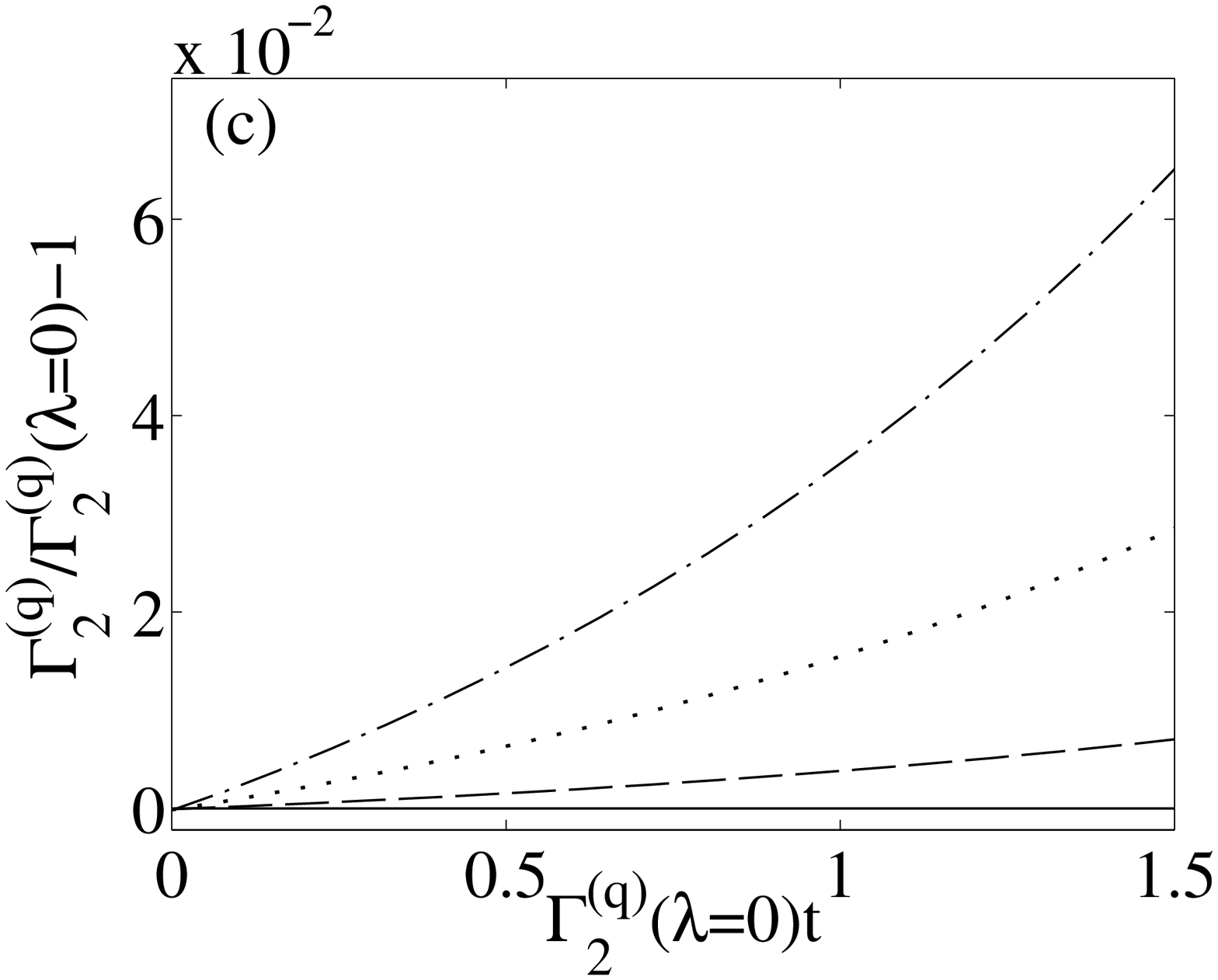}
\end{minipage}
\caption{Relative corrections to qubit averaged dephasing rate,
i.e. after eliminating fast oscillations around the slowly-varying
function, as a function of scaled time in the case of (a)
strongly, (b) moderately and (c) weakly-dissipative TLS. The ratio
$\Gamma_1^{\rm (TLS)}/\Gamma_1^{\rm (q)}$ is 10 in (a), 1.5 in (b)
and 0.1 in (c). The solid, dashed, dotted and dash-dotted lines
correspond to $\lambda/\Gamma_1^{\rm (q)}=0$, 0.3, 0.6 and 0.9,
respectively. $\theta_{\rm q} = \pi/3$ and $\theta_{\rm
TLS}=3\pi/8$.}
\end{figure*}

\subsubsection{Relaxation dynamics}

Characterizing the dynamics is most easily done by considering the
relaxation dynamics. From Fig. 1 we can see that in the case
$\lambda \neq 0$, there are several possible types of behaviour of
the qubit depending on the choice of the different parameters in
the problem. As a general simple rule, which is inspired by Fig.
1(a), we find that for small values of $\lambda$ the relaxation
rate starts at its background value and follows an exponential
decay function with a characteristic time given by $(\Gamma_2^{\rm
(TLS)}+\Gamma_2^{\rm (q)}-\Gamma_1^{\rm (q)})^{-1}$, after which
it saturates at a steady-state value given by Eq.
(\ref{eq:Pert_theory_rates}), with $E_{\rm q}=E_{\rm TLS}$:

\noindent
\begin{widetext}
\begin{equation}
\frac{dP_{\rm{ex}}(t)/dt}{P_{\rm ex}(t)-P_{\rm ex}(\infty)}
\approx - \Gamma_1^{\rm (q)} - \frac{\lambda^2 \sin^2 \theta_{\rm
q} \sin^2 \theta_{\rm TLS}}{2 \left( \Gamma_2^{\rm
(TLS)}+\Gamma_2^{\rm (q)}-\Gamma_1^{\rm (q)} \right)} \left( 1-
\exp{\left\{- \left( \Gamma_2^{\rm (TLS)}+\Gamma_2^{\rm
(q)}-\Gamma_1^{\rm (q)} \right) t\right\}} \right).
\label{eq:General_Relaxation Rate}
\end{equation}
\end{widetext}

\noindent We can therefore say that the qubit relaxation starts
with an exponential-times-Gaussian decay function for a certain
period of time, after which it is well described by an exponential
decay function with a rate that incorporates the effects of the
TLS, namely that given in Eq. (\ref{eq:Pert_theory_rates}).
Clearly the above picture is only valid when the expression for
the transient time given above is much smaller than
$1/\Gamma_1^{\rm (q)}$. Furthermore, Eq.
(\ref{eq:General_Relaxation Rate}) is not well defined when
$\Gamma_2^{\rm (TLS)}+\Gamma_2^{\rm (q)}-\Gamma_1^{\rm (q)}=0$.
However, even when the above condition about the short transient
time is not satisfied, and even when the exponent in Eq.
(\ref{eq:General_Relaxation Rate}) becomes positive, we find that
for times of the order of the qubit relaxation time, a very good
approximation for the relaxation rate is still given by Eq.
(\ref{eq:General_Relaxation Rate}). The reason why that is the
case can be seen from the expansion of Eq.
(\ref{eq:General_Relaxation Rate}) in powers of $t$:

\noindent
\begin{equation}
\frac{dP_{\rm{ex}}(t)/dt}{P_{\rm ex}(t)-P_{\rm ex}(\infty)}
\approx - \Gamma_1^{\rm (q)} - \frac{1}{2}\lambda^2 \sin^2
\theta_{\rm q} \sin^2 \theta_{\rm TLS} t,
\end{equation}

\noindent which can be integrated to give:

\noindent
\begin{equation}
P_{ex}(t) \approx \exp{\left\{-\Gamma_1^{\rm (q)} t - \lambda^2
\sin^2 \theta_{\rm q} \sin^2 \theta_{\rm TLS} t^2 /4\right\}},
\label{eq:Init_decay}
\end{equation}

\noindent where we have assumed that $P_{\rm ex}(0)=1$ and that
$P_{\rm ex}(\infty)$ is negligibly small. Eq.
(\ref{eq:Init_decay}) describes the initial decay for any ratio of
qubit and TLS decoherence times. Whether that function holds for
all relevant times or it turns into an exponential-decay function
depends on the relation between $\Gamma_1^{\rm (q)}$ and
$\Gamma_2^{\rm (TLS)}+\Gamma_2^{\rm (q)}$, as discussed above. In
particular, in the case when the TLS decoherence rates are much
smaller than those of the qubit, Eq. (\ref{eq:Init_decay}) holds
at all relevant times, and the contribution of the TLS to the
qubit relaxation dynamics is therefore a Gaussian decay function.
We also note here that the relaxation rate shows small
oscillations around the functions that we have given above.
However, those oscillations have a negligible effect when the rate
is integrated to find the function $P_{ex}(t)$.

\subsubsection{Dephasing dynamics}

The dephasing dynamics was somewhat more difficult to analyze. The
dephasing rate generally showed oscillations with frequency
$E_{\rm q}$, and the amplitude of the oscillations grew with time,
making it difficult to extract the dynamics directly from the raw
data for the dephasing rate. However, when we plotted the averaged
dephasing rate over one or two oscillation periods, as was done in
generating Fig. 2, the curves became much smoother, and we were
able to fit those curves with the following simple analytic
formula, which we obtained in an analogous manner to Eq.
(\ref{eq:General_Relaxation Rate}):

\noindent
\begin{widetext}
\begin{equation}
\frac{1}{\rho_{01}}\left(\frac{\rm{d}\rho_{01}}{\rm{d}t}\right)
\approx - \Gamma_2^{\rm (q)} -\frac{\lambda^2 \sin^2 \theta_{\rm
q} \sin^2 \theta_{\rm TLS}}{4 \left( \Gamma_2^{\rm
(TLS)}-\Gamma_2^{\rm (q)} \right)} \left( 1- \exp{\left\{- \left(
\Gamma_2^{\rm (TLS)}-\Gamma_2^{\rm (q)} \right) t\right\}}
\right). \label{eq:General_Dephasing Rate}
\end{equation}
\end{widetext}

\noindent Starting from this point, the analysis of the dephasing
dynamics is similar to that of relaxation. When the decoherence
rates of the TLS are much larger than those of the qubit, the
dephasing rate starts from its background value but quickly
reaches its steady-state value given by Eq.
(\ref{eq:Pert_theory_rates}). In the opposite limit, where the TLS
decoherence rates are much smaller than those of the qubit, a good
approximation is obtained by expanding Eq.
(\ref{eq:General_Dephasing Rate}) to first order in $t$. In that
case we find that:

\noindent
\begin{equation}
\rho_{01}(t) \approx \rho_{01}(0) \exp{\left\{-\Gamma_2^{\rm (q)}
t - \lambda^2 \sin^2 \theta_{\rm q} \sin^2 \theta_{\rm TLS} t^2
/8\right\}}.
\end{equation}

\subsection{Strong-coupling regime}

In the strong-coupling regime corresponding to large values of
$\lambda$, the qubit relaxation and dephasing rates as plotted
similarly to Figs. 1 and 2 show oscillations throughout the period
where the qubit is far enough from its thermal equilibrium state.
Therefore one cannot simply speak of a TLS contribution to qubit
decoherence. Analytic expressions can be straightforwardly derived
for the dynamics in the limit where the coupling strength is much
larger than the decoherence rates. However, the algebra is quite
cumbersome, and the results are rather uninspiring. Therefore, we
shall not present such expressions here.

\subsection{Further considerations}

\subsubsection{Comparison with traditional weak-coupling approximation}

To demonstrate the differences between our results and those of
the traditional weak-coupling approximation, we plot in Fig. 3 the
relaxation rate (at the end of the transient time) as a function
of the coupling strength $\lambda$. A similar figure can be
obtained for the dephasing rate, but we do not include it here.
Our numerical results agree with the perturbation calculation of
Sec. IV (Eq. \ref{eq:Pert_theory_rates}) up to the point where the
coupling can be classified as strong, as will be explained in Sec.
VI. Therefore we conclude that the results of our perturbation
calculation have a much wider range of validity than those of the
traditional weak-coupling approximation. The more non-negligible
the TLS decoherence times are relative to those of the qubit, the
larger the difference between the two approaches. Note that in our
perturbation-theory calculation we took the limit where $\lambda$
is much smaller than all the decoherence rates in the problem. It
turns out, however, that the results of that calculation are valid
as long as $\lambda$ is substantially smaller than the TLS
decoherence rates, assuming those are substantially larger than
the qubit decoherence rates. No specific relation is required
between $\lambda$ and the qubit decoherence rates in that case.

\begin{figure}[ht]
\includegraphics[width=8.5cm]{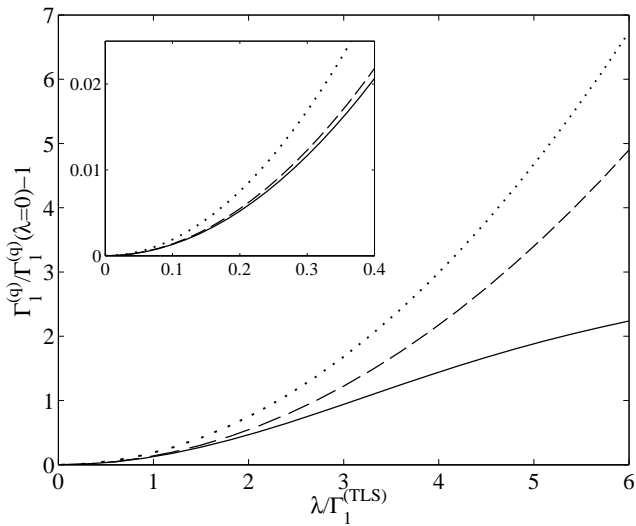}
\caption{\label{fig3b} The steady-state qubit relaxation rate as a
function of the qubit-TLS coupling strength $\lambda$. The solid
line represents the numerical results, the dashed line is the
perturbation-calculation result of Sec. IV and the dotted line is
the result of the traditional weak-coupling approximation given by
Eq. (\ref{eq:Semiclassical_rates}). $\theta_{\rm q} = 3\pi/20$,
$\theta_{\rm TLS}=2\pi/5$, $\Gamma_1^{\rm (q)}/\Gamma_1^{\rm
(TLS)}=0.25$, $\Gamma_2^{\rm (q)}/\Gamma_1^{\rm (TLS)}=1$, and
$\Gamma_2^{\rm (TLS)}/\Gamma_1^{\rm (TLS)}=2$.}
\end{figure}

For further demonstration of the differences between the
predictions of the two approaches, we ran simulations of an
experiment where one would sweep the qubit energy splitting and
measure the relaxation and dephasing rates. We used a TLS with the
parameters of Fig. 3(a) and $\lambda$ ranging from 0.15 to 0.5. In
such an experiment, one would see a peak in the relaxation and
dephasing rates at the TLS energy splitting. According to the
traditional weak-coupling approximation, the height of the
dephasing peak should be half of that of the relaxation peak. In
the numerical simulations with the above parameters, we see a
deviation from that prediction by about 25\%. The relation between
the shapes of the two peaks agrees very well with the expressions
in Eq. (\ref{eq:Pert_theory_rates}). That difference would, in
principle, be measurable experimentally. Note, however, that since
we are dealing with an uncontrollable environment, there is no
guarantee that a TLS with the appropriate parameters will be found
in the small number of qubit samples available at a given
laboratory.

\subsubsection{Two TLSs}

In order to establish that our results are not particular to
single quantum TLSs, we also considered the case of two TLSs that
are both weakly coupled to the qubit. If we take the two TLS
energy splittings to be larger than the widths of their
frequency-domain correlation functions, we find that the
relaxation dynamics is affected by at most one TLS, depending on
the qubit energy splitting. The TLS contributions to pure
dephasing dynamics, i.e. that unrelated to relaxation, are
additive, since that rate depends on the zero-frequency noise. We
then considered two TLSs with energy splittings equal to that of
the qubit. We found that the TLS contributions to the qubit
relaxation and dephasing dynamics are additive in both the large
and small $\Gamma^{\rm (q)}/\Gamma^{\rm (TLS)}$ limits. These
results are in agreement with those found in Ref.
\cite{Fedichkin2}, where a related problem was treated.

\subsubsection{Entanglement}

It is worth taking a moment here to discuss the question of
entanglement between the system and environment. It is commonly
said that in a Markovian master equation approach the entanglement
between a system and its surrounding environment is neglected, a
statement that can be misinterpreted rather easily. In order to
address that point, we consider the following situation: we take
the parameters to be in the weak-coupling regime, where such a
discussion is meaningful. We take the qubit to be initially in its
excited state, with no entanglement between the qubit and the TLS.
We find that the off-diagonal matrix elements of the combined
system density matrix describing coherence between the states
$\ket{\uparrow_{\rm q} \downarrow_{\rm TLS}}$ and
$\ket{\downarrow_{\rm q} \uparrow_{\rm TLS}}$ start from zero at
$t=0$ and reach a steady state at the end of the transient time.
Beyond that point in time, they decay with the same rate as the
excited state population. We therefore conclude that the final
relaxation rate that we obtain takes into account the effects of
entanglement between qubit and TLS, even though the density matrix
of the qubit alone exhibits exponential decay behaviour.

\section{Criteria for strong coupling between qubit and two-level system}

There are a number of possible ways one can define the criteria
distinguishing between the weak and strong coupling regimes. For
example, one can define a strongly-coupled TLS as being one that
contributes a decoherence rate substantially different from that
given by some weak-coupling analytic expression. One could also
define a strongly-coupled TLS as being one that causes visible
oscillations in the qubit dynamics, i.e. one that causes the
relaxation and dephasing rates to change sign as time goes by. We
shall use the criterion of visible deviations in the qubit
dynamics from exponential decay as a measure of how strongly
coupled a TLS is.

Even with the above-mentioned criterion of visible deviations in
the qubit dynamics from exponential decay, one still has to
specify what is meant by visible deviations, e.g. maximum
single-point deviation or average value of deviation. One also has
to decide whether to use relaxation or dephasing dynamics in that
definition. We have used a number of different combinations of the
above and found qualitatively similar results. Those results can
essentially be summarized as follows: a given TLS can be
considered to interact weakly with the qubit if the coupling
strength $\lambda$ is smaller than the largest (background)
decoherence rate in the problem. The exact location of the
boundary, however, varies by up to an order of magnitude depending
on which part of the dynamics we consider and how large a
deviation from exponential decay we require.

We have also checked the boundary beyond which our analytic
expressions and numerical results disagree, and we found that the
boundary is similar to the one given above. That result confirms
the wide range of validity of our analytic expressions. Note in
particular that even if the qubit-TLS coupling strength $\lambda$
is larger than the decoherence rates of the qubit, that TLS can
still be considered weakly coupled to the qubit, provided the TLS
decoherence rates are larger than $\lambda$.

\section{Conclusion}

We have analyzed the problem of a qubit interacting with a quantum
TLS in addition to its coupling to a background environment. We
have characterized the effect of the TLS on the qubit decoherence
dynamics for weak and strong coupling, as well as weakly and
strongly-dissipative TLSs. We have found analytic expressions for
the contribution of a single TLS to the total decoherence rates in
the weak-coupling regimes, which is a much larger range than just
the weak-coupling limits. We recover the results of the
traditional weak-coupling approximation as a special case of our
results, namely for a weakly-coupled strongly-dissipative TLS. We
have found that weakly-coupled weakly-dissipative TLSs exhibit
memory effects by contributing a non-exponential factor to the
qubit decoherence dynamics. We have verified that the
contributions of two TLSs to the qubit relaxation and dephasing
rates are additive in the weak-coupling limit. We have discussed
the transition from weak to strong coupling and numerically found
that the transition occurs when the qubit-TLS coupling strength
exceeds all the decoherence rates in the problem.

\begin{acknowledgments}
This work was supported in part by the National Security Agency
(NSA) and Advanced Research and Development Activity (ARDA) under
Air Force Office of Research (AFOSR) contract number
F49620-02-1-0334; and also supported by the National Science
Foundation grant No.~EIA-0130383. One of us (S. A.) was supported
in part by a fellowship from the Japan Society for the Promotion
of Science (JSPS).
\end{acknowledgments}


\begin{thebibliography}{99}
\bibitem{Bloch_sphere} In this representation, a given $2 \times 2$
density matrix $\rho$ is mapped onto the point $\vec{P}$ that
satifies the equation $\rho = \frac{1}{2} ( 1 + \vec{P} \cdot
\vec{\sigma} )$, where $\vec{\sigma}$ is the vector of Pauli
matrices.
\end{thebibliography}
\end{document}